\documentclass[12pt]{iopart}

\usepackage{graphicx}
\usepackage{dcolumn}
\usepackage{bm}
\usepackage{braket}
\usepackage[usenames,dvipsnames]{xcolor}
\expandafter\let\csname equation*\endcsname\relax

\expandafter\let\csname endequation*\endcsname\relax

\usepackage{amsmath,amsthm}
\graphicspath{{./Figure/}{./}}

\newcommand{\im}{\ensuremath{\mathrm{i}}} 

\def\bra#1{\mathinner{\langle{#1}|}}
\def\ket#1{\mathinner{|{#1}\rangle}}

\newcommand{\ave}[1]{{\langle #1\rangle}}

\newcommand{\HH}{\mathcal{H}}
\newcommand{\BBH}{\mathcal{B}(\mathcal{H})}

\def\bra#1{\mathinner{\langle{#1}|}}
\def\ket#1{\mathinner{|{#1}\rangle}}

\newtheorem{thm}{Theorem}

\newtheorem{remark}{Remark}

\newcommand{\LL}{{\hat{\cal L}}}

\def\tr{{{\rm tr}}}

\definecolor{OxfBlue}{rgb}{0, 0.333, 0.710}

\begin{document}

\title[Stationary state degeneracy of open quantum systems
with non-Abelian symmetries]{Stationary State Degeneracy of Open Quantum Systems with Non-Abelian Symmetries}

\author{Zh. Zhang$^{1,2}$, J. Tindall$^{1}$, J. Mur-Petit$^{1}$, D. Jaksch$^{1}$ and B. Bu\v{c}a$^{1}$}
\address{$^1$Clarendon Laboratory, University of Oxford, Parks Road, Oxford OX1 3PU, United Kingdom}
\address{$^2$Department of Physics, Tsinghua University, Beijing 100084, P. R. China}

\date{\today}

\begin{abstract}
We study the null space degeneracy of open quantum systems with multiple non-Abelian, strong symmetries. By decomposing the Hilbert space representation of these symmetries into an irreducible representation involving the direct sum of multiple, commuting, invariant subspaces we derive a tight lower bound for the stationary state degeneracy. We apply these results within the context of open quantum many-body systems, presenting three illustrative examples: a fully-connected quantum network, the XXX Heisenberg model and the Hubbard model. We find that the derived bound, which scales at least cubically in the system size the $SU(2)$ symmetric cases, is often saturated. Moreover, our work provides a theory for the systematic block-decomposition of a Liouvillian with non-Abelian symmetries, reducing the computational difficulty involved in diagonalising these objects and exposing a natural, physical structure to the steady states - which we observe in our examples.
\end{abstract}


\submitto{\jpa}

\maketitle

\section{Introduction}
Understanding ergodicity in quantum many-body systems remains a fundamental task of mathematical physics. The notion of symmetries plays a crucial role in determining the dynamics and long-time behaviour  \cite{ETHReview,VidmarRigol,GETH,ChargesReview,doyon,ChiralBICs,vanHoveBICs}. For closed many-body systems the existence of extensive symmetries (e.g. due to the underlying integrability, or localization of the model) can lead to the absence of both thermalization and ergodicity \cite{EsslerGGE,Pozsgay2,EnejGGE,quench1,quench5,XXZ1,XXZ2,quench2}. 

Understanding non-ergodicity of open quantum systems is seemingly more difficult. For finite-size systems within the Markovian approximation, i.e. when the time dynamics can be described in the Lindblad master equation framework \cite{Lindblad,GKS,openbook}, the question is partially resolved \cite{BaumgartnerNarnhofer,Buca2012,AlbertJiang1,Albert2}. It is known that specific kinds of symmetries called \emph{strong symmetries} \cite{Buca2012} lead to degeneracy of the (time-independent) stationary state to which the open system evolves into in the long-time limit. This result connecting the structure of the symmetry operator and the degeneracy of the stationary state has allowed research into many interesting questions, e.g. quantum memory storage and manipulation \cite{AlbertJiang2}, quantum metrology \cite{metrology}, quantum batteries \cite{battery}, quantum transport \cite{HurtadoManzano1,Manzano2}, exact solutions \cite{Enej}, potential probes of molecular structure \cite{Thingna2016}, dissipative state preparation \cite{Znidaric,Igor}, correlation functions \cite{Kos} and thermalization and relaxation \cite{Gritsev,Wolff,Carr}. Furthermore, degeneracy of the stationary states can be a starting point for realizing discrete time crystals under dissipation \cite{Lazarides,JuanDTC,2019time}. 

The strong symmetry theorem \cite{Buca2012} has thus far been limited to Abelian symmetries. In this article we derive a lower bound for the stationary-state degeneracy of an open quantum system with multiple, non-Abelian strong symmetries. We achieve this through decomposing the Hilbert space representation of these symmetries into a series of irreducible representations. As illustrative examples, we apply our theory to three archetypal models: a quantum network, the open randomly dissipative XXX Heisenberg model and the spin-dephased Fermi-Hubbard model. The models are interesting for understanding transport in molecules and quantum networks \cite{HurtadoManzano1,thingna2019magnetic}, effects of disorder on relaxation, and superconductivity out-of-equilibrium \cite{Tindall2019}, respectively. We also study the XXX model under collective dissipation as a simple toy example. Using our theory we are able to derive a tight bound for the stationary state degeneracy in each case, finding numerically that this bound is saturated in most situations.

The stationary-state degeneracy in the XXX spin chain and Hubbard model examples, which scales at least cubically in the system size, provides a space for the storage of quantum information -- despite the presence of environmental noise. Moreover, our work provides a theory for the systematic block-decomposition of a Liouvillian with non-Abelian symmetries, reducing the computational difficulty involved in diagonalising these objects and exposing a natural, physical structure to the steady states which we observe in our examples.

\section{Theoretical lower bounds on the stationary state degeneracy of an open quantum system}

Under the Markov approximation \cite{Lindblad,GKS,GardinerZoller, openbook}, the dynamical evolution of an open quantum system can be described by the Lindblad master equation (here, and in the remainder of this work, we set $\hbar = 1$)
\begin{equation}
    \dot{\rho}=\mathcal{L}\rho
    =-\im[H,\rho]+\sum_{l}\gamma_l(2L_l{\rho}L_l^{\dag}
     -\{L^{\dag}_l L_l,\rho\}),
\label{Eq: LindbladEqn}
\end{equation}
where $\rho$ is the density matrix of the system, \(\mathcal{L}\) is the Liouvillian superoperator acting on the space of bounded linear operators $\BBH$. The Hamitonian $H$ in Eq. ~(\ref{Eq: LindbladEqn}) describes the coherent evolution of the system while the Lindblad `jump' operators $L_{l}$ describe the interactions between the system and the environment with corresponding coupling strengths $\gamma_{l}$.

We also define the adjoint of the master equation via the standard Heisenberg picture for the observable $O$ $\ave{O(t)}=\tr\left({O \exp(\LL t)\rho(0)}\right)=\tr\left({\exp(\LL^\dagger t)O \rho(0)}\right)$ (expanding the exponential),
\begin{equation}
    \dot{O}=\mathcal{L}^\dagger O
    =\im[H,\rho]+\sum_{l}\gamma_l(2L^{\dag}_l{\rho}L_l
     -\{L^{\dag}_l L_l,\rho\}),
\label{Eq: LindbladEqnAdjoint}
\end{equation}
which also defines the adjoint Liouvillian superator $\LL^\dagger$. 
\subsection{Stationary state degeneracy in the presence of Abelian strong symmetries}

As a starting point for our theory we describe the nullspace degeneracy of an open quantum system in the presence of a single strong symmetry \cite{Buca2012}. A strong symmetry $S$ of the Lindblad equation is a unitary operator satisfying the commutation relations
\begin{equation}
    [H,S]=0 , \quad [L_l,S]=[L_l^\dag,S]=0, \quad \forall l. \label{Eq:Strong}
\end{equation}
Now suppose $S$ has $n_{s}$ different eigenvalues. Then we can decompose the Hilbert space into the corresponding blocks for each eigenvalue
\begin{equation}
    \mathcal{H}=\bigoplus_{\alpha = 1}^{n_s}\mathcal{H}_\alpha.
\end{equation}

Moreover, we can carry this decomposition through to the Banach space\footnote{The Banach space is created by performing a thermo-field doubling and mapping matrix elements $\ket{\psi}\bra{\phi} \to \ket{\psi}\otimes \ket{\phi}$. This `vectorization' allows for a natural Hilbert-Schmidt inner product to be defined on $\BBH$, $(A,B):=\tr A^\dagger B$ with $A,B\in \BBH$.} of bounded linear operators \(\BBH =(\mathcal{H},\mathcal{H})\)
\begin{equation}
    (\mathcal{H},\mathcal{H})=\bigoplus_{\alpha=1}^{n_s}\bigoplus_{\beta=1}^{n_s}(\mathcal{H}_\alpha,\mathcal{H}_\beta),
\end{equation}
where the Liouvillian is block-diagonal, i.e. $\mathcal{L}$ is invariant in each subspace $(\mathcal{H}_\alpha,\mathcal{H}_\beta)$
\begin{equation}
  \mathcal{L}(\mathcal{H}_\alpha,\mathcal{H}_\beta)
  \subseteq (\mathcal{H}_\alpha,\mathcal{H}_\beta),
\end{equation}
which follows from \eqref{Eq:Strong}. 
Furthermore, it can be proven, due to trace preservation of the dynamics, that in every diagonal subspace $(\mathcal{H}_\alpha,\mathcal{H}_\alpha)$ there is at least 1 stationary state of $\mathcal{L}$ \cite{Buca2012}. Hence, the stationary-state degeneracy of a system with a single strong symmetry has a lower bound of $n_s$. 

\subsection{Lower bound on the null space dimension for systems with non-Abelian strong symmetries}\label{Sec:LowerBoundTheory}
We now prove the main result of this article: the lower bound of the null space in the case of multiple, non-Abelian strong symmetries. We start by stating the following simple, but useful, theorem.  


\begin{thm} \label{theorem}
Let the dynamics of a system be given by a Lindblad master equation, i.e. equation \eqref{Eq: LindbladEqn}, with the Hamiltonian $H \in \BBH$ and the set of Lindblad jump operators $L_l \in \BBH$ being bounded linear operators acting over a Hilbert space $\HH$. Let there exist a set $\{S_1, S_2, ...,S_n\}$ of \emph{strong} symmetries which form a non-Abelian group,
\begin{equation}
 [H,R(S_k)]=0, \quad
 [L_l,R(S_k)]=[L_l^\dag,R(S_k)]=0,\quad \forall l,k,
 \label{eq:RTk-commut}
\end{equation}
where $R(S_{k})$ is the representation of the $kth$ strong symmetry in the Banach space\footnote{In general, due to these symmetries being non-Abelian, we have that $[R(S_{k}), R(S_{k'})] \neq 0$.}. We then perform a decomposition of the group representation $R$ into a series of irreducible representations, 
\begin{equation}
 R = {\bigoplus}_{i=1}^s R_i,
\end{equation}
with each representation $R_i$ having a dimension of $D_i$ (i.e. it is a matrix of size $D_i \times D_i$). The degeneracy (dimension) of the stationary state $\LL \rho_\infty=0$ is then bounded from below by $\sum_{i=1}^s D_i^2$. 
\end{thm}
\begin{proof}
Define $\Tilde{R}_i$ to be an operator on the full Hilbert space acting as $R_i$ in the irrep subspace and trivially in the the rest of the Hilbert space. Following the decomposition we have, $[H,\Tilde{R}_i(S_k)]=[L_{l},\Tilde{R}_i(S_k)]=0, \forall i,k,l,$ and by Schur's representation lemma this implies that $H,L_l,L_l^\dagger$ are multiplies of the identity matrix in the $R_i$ blocks  \cite{group}. This, in turn, implies that
\begin{equation}
[H,\{R_i(S_k)\}_{nm}]=[L_l,\{\Tilde{R}_i(S_k)\}_{nm}]=[L_l^\dagger,\{\Tilde{R}_i(S_k)\}_{nm}]=0,    
\end{equation} 
where $n$ and $m$ index the elements of the matrix $\Tilde{R}_i(S_k)$. This means that there are $\sum_{i=1}^s D_i^2$ linearly independent operators that commute with $H,L_l,L_l^\dagger$. It immediately follows that these operators are in the kernel of $\LL^\dagger$, which has the same dimension as the kernel of $\LL$.   
\end{proof}

\begin{remark}
Note that the individual states in the representation $\ket{i,j}$, with $i=1,\ldots,s$ indexing the irreducible representation and $j=1,\ldots,D_i$ indexing the states inside the irreducible representation, are often highly degenerate. By Schur's lemma both $H$ and $L_l$ are proportional to the identity matrix on the subspace $\ket{i,j}$. However, in general they can have further symmetry structure within the $\ket{i,j}$ subspace (e.g. they can also be multiples of the identity matrix inside these subspaces). In those cases the degeneracy will be larger than the bound; we explore an example of this situation in Sec.~\ref{ssec:total-spin-dissip}. In many situations, however, we find there is no additional structure and that this bound is saturated. \label{rem1} 
\end{remark}

\begin{remark}
Assuming standard unitary strong symmetries $S_k$ it is easy to see that $[L_l,S_k]=0$ implies $[L^\dagger_l,S_k]=0$. In that case we may dispense with this final requirement of Th.~\ref{theorem}. 
\end{remark}

\begin{remark}
Due to the non-Abelian nature of the symmetry group there may exist stationary states of $\LL \rho_{\infty,k}=0$ that are not density matrices, i.e. $\rho^\dagger_{\infty,k} \neq \rho_{\infty,k}$. However, by nature of the dynamical map in equation \eqref{Eq: LindbladEqn} every initial density matrix $\rho(0)$ will always remain a valid density matrix and states of the form $\rho^\dagger_{\infty,k} \neq \rho_{\infty,k}$ are components for valid density matrices in the long-time limit. These states are interesting from a quantum information and computing perspective because they represent quantum coherences that can be used to store quantum information \cite{MultipleSteadyStates}.  
\end{remark}

In the following, we illustrate these results with a series of examples.

\section{Examples}

Now that we have identified the lower bound in Theorem ~\ref{theorem}, we apply this result to several qualitative different models: a fully-connected quantum network, the Heisenberg model, and the Hubbard model. Physically, the first model is used to study energy transport in molecules, the second for studying quantum magnetism and the the third one for studying strongly correlated electrons. 

In each example, we use exact diagonalization to compute the steady state degeneracy numerically using the QuTiP library~\cite{qutip2012,qutip2013}, and compare the numerical results with the lower bound in Theorem ~\ref{theorem}. 

\subsection{Fully-connected quantum network}
\begin{figure}[t]
\begin{indented}
 \item   \includegraphics[width=0.8\columnwidth]{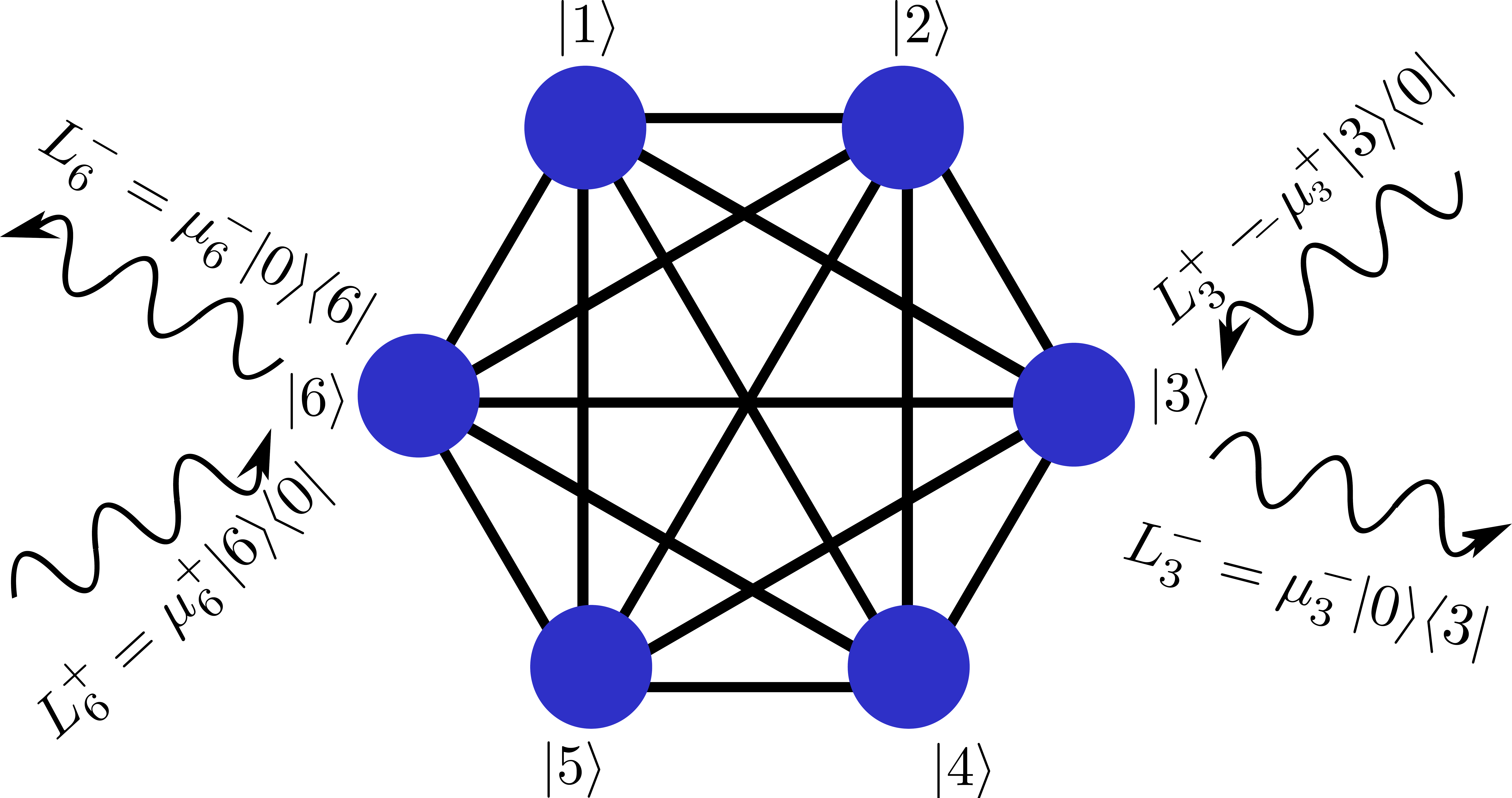}
\end{indented}
 \caption{Sketch of the fully-connected network described in the text for the particular case of $N = 6$ sites. Lines represent coherent hopping between sites. The wavy arrows represent absorption and emission of excitons between the system and the environment.}
\label{Fig:Network}
\end{figure}

As our first example, we consider a fully-connected quantum network used as a simple model for studying the role of symmetries in energy transport in molecules~\cite{Thingna2016,HurtadoManzano2}. The model describes an exciton hopping between a series of interconnected lattice sites, pictured in figure~(\ref{Fig:Network}); similar lattice models are frequently used to represent the dynamics of atoms in optical lattices, electrons in quantum-dot arrays, and photons in photonic waveguides or quantum circuits~\cite{Victor}. These types of models have also proven useful in studying excitonic transport in photosynthetic light harvesting systems \cite{cao2009optimization,Buchleitner}. 

The Hilbert space is spanned by the basis \(\{\vert 0{\rangle}\), \({\mid}1{\rangle}\), \({\mid}2{\rangle}\),..., \({\mid}N{\rangle}\}\) where \({\mid}i{\rangle}\) denotes the state with an exciton on site \(i=1,\ldots,N\) while \({\mid}0{\rangle}\) is the ground state characterised by the absence of any excitation.
In this basis, the Hamiltonian of the system reads
\begin{eqnarray}
H
& = \varepsilon_{g}{\mid}0{\rangle}{\langle}0{\mid}
 + \varepsilon\sum_{i=1}^{N}{{\mid}i{\rangle}{\langle}i{\mid}}
 + h\sum_{i\neq j}{{\mid}i{\rangle}{\langle}j{\mid}}
=
 \left[ \begin{array}{ccccc}
   \varepsilon_{g} & 0 & 0 & ... & 0\\
   0 & \varepsilon & h & ... & h\\
   0 & h & \varepsilon & ... & h\\
   \vdots & \vdots & \vdots & \ddots & \vdots\\
   0 & h & h & ... & \varepsilon
   \end{array} \right] \:,
\end{eqnarray}
where \(\varepsilon_{g}\), \(\varepsilon\) and \(h\) play the role of ground state, excitation and kinetic energy scales respectively. 

In order to drive a current through the network some of the lattice sites are connected to thermal baths which each interact with the system via the jump operators
\begin{eqnarray}
L_j^+=\mu_j^+{\mid}j{\rangle}{\langle}0{\mid}\nonumber\\
L_j^-=\mu_j^-{\mid}0{\rangle}{\langle}j{\mid}.
\end{eqnarray}
These operators describe the absorption and emission of an exciton on site \(j\) where \(\mu_j^{\pm}\) is the relevant probability of that process.

The Liouvillian is invariant under the permutation of any two sites which are not connected to the thermal baths. This symmetry can be described by the exchange operator for sites \(i\) and \(j\)
\begin{equation}
   P_{ij}=I-{\mid}i{\rangle}{\langle}i{\mid}-{\mid}j{\rangle}{\langle}j{\mid}+{\mid}i{\rangle}{\langle}j{\mid}+{\mid}j{\rangle}{\langle}i{\mid}.
\end{equation}
The exchange operators that don't act on the bath sites (we assume label the bath sites as $a, b, ...$) then form a series of strong symmetries:
\begin{equation}
[H,P_{ij}]=0, \quad [L^{\pm}_{a},P_{ij}] = [L^{\pm}_{b},P_{ij}]= \hdots = 0,\quad \forall i, j \neq a,b, ....
\end{equation}
Moreover, these operators do not necessarily commute $[P_{i,j}, P_{k,l}] \neq 0 \ \forall i, j, k ,l$ and thus constitute a set of non-Abelian strong symmetries in this open quantum system.

More generally, every group element $A$ in the permutation group  \(P\), which is formed from the different products of exchange elements
\begin{equation}
    A=\prod_{i, j}P_{ij},
\end{equation}
is a strong symmetry, i.e.,
\begin{equation}
    [H,A]=0, \quad [L_l,A]=0,\quad\forall l.
\end{equation}
These group elements are thus left null eigenvectors of the Liouvillian~\cite{Buca2012,AlbertJiang1}
\begin{equation}
    \mathcal{L}^\dag(A)=0.
\end{equation}

Now suppose we label the sites which are uncoupled form the baths as $1,2, ..., n$. An arbitrary permutation operator can be block diagonalized as follows
\begin{equation}
 A =
 \left[ \begin{array}{cc}
   R_n(A)  & 0\\
   0 & I_{N-n}
   \end{array} \right],
\end{equation}
where $R_{n}(A)$ is the natural representation of \(A\), and acts on the space spanned by \({\mid}1{\rangle}\), \({\mid}2{\rangle}\),...,\({\mid}n{\rangle}\). Meanwhile, $I_{N-n}$ acts on the remaining space, spanned by \({\mid}0{\rangle}\), \({\mid}n+1{\rangle}\),\({\mid}n+2{\rangle}\),...,\({\mid}N{\rangle}\). We anticipate this block-diagonal structure will emerge in the stationary states of the model. Moreover, the natural representation is reducible as it can be written as the direct sum of the irreducible, trivial and standard representations \cite{Fulton2004}
\begin{equation}
    R_n(A)=R_t(A)\oplus R_s(A),
    \label{Eq:TrivialNatural}
\end{equation}
whose ranks are 1 and $n-1$, respectively.
Hence, the permutation operator \(P\) is further block diagonalized:
\begin{equation}
 A=
 \left[ \begin{array}{ccccc}
   1 & 0 & 0\\
   0 & R_s(P) & 0\\
   0 & 0 & I_{N-n}\\
   \end{array} \right].
\label{Eq:BlockDiagonalPerm}
\end{equation}

We have thus found an irreducible representation of the strong symmetries and since the left kernel of \(\mathcal{L}\) contains all possible \(A\), by Theorem ~\ref{theorem},  its dimension must be at least \((n-1)^2+1\). Furthermore, for every left null vector of \(\mathcal{L}\), there is a corresponding right null vector and the dimension of the right kernel is equal to the left kernel dimension. Thus the stationary state degeneracy of this model is at least \((n-1)^2+1\).

In table~\ref{table:network} we compare this result to those obtained via exact diagonalization of the full Liouvillian. The results are in complete agreement for all the system sizes we are able to reach and the bound is always completely saturated.

\begin{table}
 \caption{\label{table:network} Steady state degeneracy of the dissipative quantum network pictured in figure~\ref{Fig:Network} for various system sizes, $N$. Two sites are always coupled to the baths. Predicted degeneracy is calculated using Theorem ~\ref{theorem} whilst the actual degeneracy is calculated by exact diagonalization of the matrix representation of the Liouvillian superoperator.}
 \begin{indented}
\item \begin{tabular}{ccc} 
  $N$ & Degeneracy lower bound & Actual degeneracy \\ 
  \mr 
  3 & 1 & 1 \\ 
  5 & 5 & 5 \\ 
  10 & 50 & 50 \\ 
  20 & 290 & 290 \\ 
  50 & 2210 & 2210 \\
  \br 
  \end{tabular}
\end{indented}
\end{table}

Additionally, we can explicitly show the decomposition in equation~(\ref{Eq:TrivialNatural}). The non-trivial part of \(A\) acts on the space spanned by \(\{{\mid}1{\rangle}\), \({\mid}2{\rangle}\),...,\({\mid}n{\rangle}\}\). To explicitly uncover the decomposition in equation~(\ref{Eq:TrivialNatural}) we change from this configuration basis to a new one:
\begin{equation}
\{\ket{1}, \ket{2}, ..., \ket{n}\} \rightarrow \{\ket{\phi_{t}}, \ket{\psi_{1}}, ..., \ket{\psi_{n-1}}\} ,
\end{equation}
where \({\mid}\phi_t{\rangle}\) is the trivial representation and is defined as
\begin{equation}
   {\mid}\phi_t\rangle=\frac{1}{\sqrt{n}}\sum_{i=1}^n{\mid}i\rangle ,
\end{equation}
which satisfies $R_{n}(A)\ket{\phi_{t}} = \ket{\phi_{t}}$. 
Meanwhile the \({\mid}\psi_i{\rangle}\) form the standard representation $R_{s}(P)$ and describe $n-1$ linearly independent basis vectors whose coefficients in this same basis sum to $0$ and thus $\braket{\phi_{t}|\psi_{i}} = 0 \ \forall i$. Note that the exciton is fully delocalised over the sites uncoupled from the baths in all states in this new basis.
\par
\begin{figure}[tb]
\begin{indented}
 \item  \includegraphics[width=0.8\columnwidth]{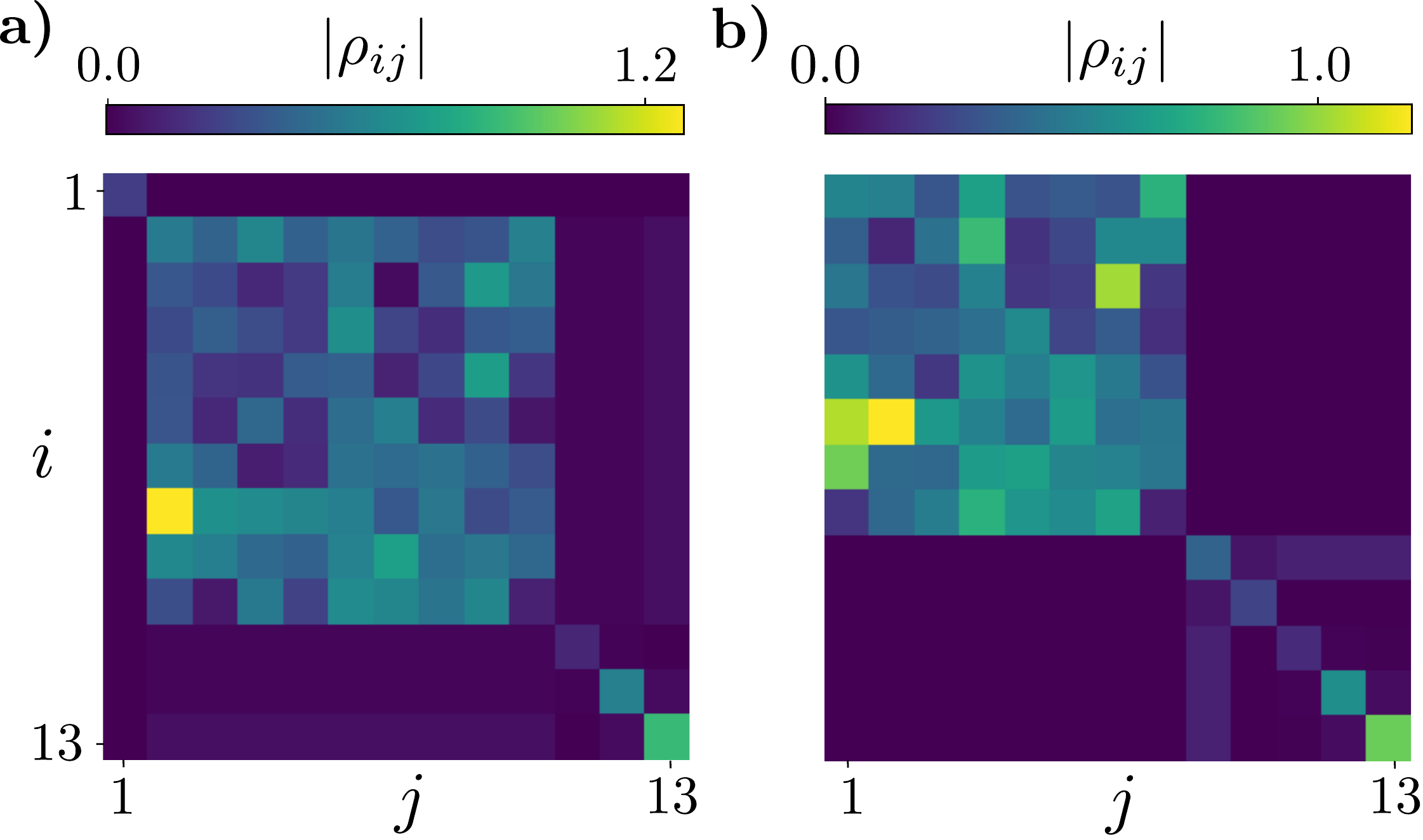}
\end{indented}
 \caption{\label{Fig:SteadyStateStructure}
 Plots of the matrix structure of the stationary states, with $i$ and $j$ indexing the matrix elements (a), (b) Example stationary state of a $N=12$ site system.
 Parameters are: \(\varepsilon_g=0\), \(\varepsilon=10\) and \(h=20\).
 Three sites are coupled to thermal baths with coupling strengths  \(\mu_{10}^-=0.6\), \(\mu_{10}^+=0.3\), \(\mu_{11}^-=0.2\), \(\mu_{11}^+=0.5\), \(\mu_{12}^-=0.2\), and \(\mu_{12}^+=0.8\).
 (a) Absolute magnitude of the matrix elements of one of the stationary states in the configuration basis where $i$ and $j$ run over the basis states $\{\ket{0}, \ket{1}, ..., \ket{12}\}$.
 (b) Absolute magnitude of the matrix elements of the same stationary state but in the basis corresponding to representations of permutation group where $i$ and $j$ run over the basis states $\{\ket{\psi_{1}}, \ket{\psi_{2}}, \ldots, \ket{\psi_{n}}, \ket{\phi_{t}}, \ket{0}, \ket{10}, \ket{11}, \ket{12}\}$.}
\end{figure}

 In figure~\ref{Fig:SteadyStateStructure} we provide example plots of the stationary states of the system following exact diagonalization. The block-diagonal structure of the steady states in the two bases exposes the decompositions in equations~(\ref{Eq:TrivialNatural}) and ~(\ref{Eq:BlockDiagonalPerm}), which provide a natural structure to the steady states of the system.

Finally, we note that the Hamiltonian and every Lindblad operators can be likewise block-diagonalized within this decomposition, and take the form:
\begin{eqnarray}
 H=
 \left[ \begin{array}{ccccc}
   \varepsilon+(n-1)h & 0 & h^\dag\\
   0 & (\varepsilon-h)I_{n-1} & 0\\
   h & 0 & H_o
   \end{array} \right] ,
 \quad
 L_i^{\pm}=
 \left[ \begin{array}{ccccc}
   0 & 0 & 0\\
   0 & 0 & 0\\
   0 & 0 & L_{i,o}^{\pm}\\
   \end{array} \right].
\end{eqnarray}
Here, we have set, without loss of generality, the ground state energy \(\varepsilon_g\) to zero.


\subsection{XXX Heisenberg model\label{sec:xxx-heisenberg}}

For our second example we turn our attention to a many-body system with continuous non-Abelian symmetries: the quantum Heisenberg model with zero magnetic field. However, most of the following discussions will be valid for general $SU(2)$ symmetric Hamiltonians.  We consider the isotropic XXX spin chain in which the spin-spin couplings are the same in all directions
\begin{equation}
H=\sum_{i=1}^{N-1}(\sigma_i^x\sigma_{i+1}^x+\sigma_i^y\sigma_{i+1}^y+\sigma_i^z\sigma_{i+1}^z),
\end{equation}
where \(\sigma_i^{x,y,z}\) denotes the $x,y,z$ Pauli operator on site \(i\).
Importantly, the system has an $SU(2)$ symmetry, which can be encovered through the total spin operators
\begin{equation}
 [H,S^x]=[H,S^y]=[H,S^z]=0,    
 \label{Eq:SpinSU2}
\end{equation}
where \(S^{x,y,z}=\sum_{i=1}^N\sigma_i^{x,y,z}\) is the total spin operator in the corresponding direction.

We use \(\textbf{2}\) to denote the fundamental representation of the SU(2) symmetry on a single site, where, more generally, \(\mathbf{n}\) denotes a representation of this Lie algebra over an \(n\) dimensional space \cite{Georgi2018}. We then use this notation to decompose the representation of the SU(2) symmetry over an $N$-site system into a series of irreducible representations. Assuming that the dissipation does not break the SU(2) symmetry of the model, the lower bound of stationary state degeneracy can be determined using Theorem ~\ref{theorem}.
In table~\ref{table:XXX} we depict this, showing the decomposition of the spin \({\rm SU(2)}\) symmetry over an
$N$-site
system and the corresponding lower bound on the stationary state degeneracy, which grows as $O(N^{3})$

\begin{table}
 \caption{\label{table:XXX}
 Decomposition of the SU(2) symmetry of the $N$-site XXX Heisenberg model into a series of fundamental, irreducible representations. For example, \(\textbf{2}^{\otimes 2}=\textbf{3}\oplus\textbf{1}\) refers to the decomposition of two spin 1/2s into a spin singlet and triplet. The third column is the corresponding lower bound for the stationary state degeneracy of the open XXX model, provided the dissipation preserves the overall SU(2) symmetry.}
 \begin{indented}
 \item
   \begin{tabular}{clr} 
   \br 
   $N$ & Decomposition of representation & Degeneracy lower bound
   \\ \mr 
    2 
      & $\textbf{2}^{\otimes 2}=\textbf{3}\oplus\textbf{1}$ 
      & $3^2+1^2=10$
      \\ \mr 
    3
      & $\textbf{2}^{\otimes 3}=\textbf{4}\oplus (2\times\textbf{2})$
      & $4^2+2^2=20$
      \\ \mr 
    4
      & $\textbf{2}^{\otimes 4}=\textbf{5}\oplus (3\times\textbf{3}) \oplus (2\times\textbf{1})$
      & $5^2+3^2+1^2=35$
      \\ \mr 
    5
      & $\textbf{2}^{\otimes 5}=\textbf{6}\oplus (4\times\textbf{4}) \oplus (5\times\textbf{2})$
      & $6^2+4^2+2^2=56$
      \\ \mr 
    6
      & $\textbf{2}^{\otimes 6}=\textbf{7}\oplus (5\times\textbf{5}) \oplus (9\times\textbf{3})\oplus (5\times\textbf{1})$
      & $7^2+5^2+3^2+1^2=84$
      \\ \mr 
    7
      & $\textbf{2}^{\otimes 7}=\textbf{8}\oplus (6\times\textbf{6}) \oplus (14\times\textbf{4})\oplus (14\times\textbf{2})$
      & $8^2+6^2+4^2+2^2=120$
      \\ \br 
 \end{tabular}
 \end{indented}
\end{table}

In order to demonstrate the strength of this bound, we consider coupling the system to two different SU(2) preserving environments.

\subsubsection{Random coupling dissipation.}
In the first case we couple the system to an environment which induces dissipation between arbitrary pairs of sites with random (complex) strength. Such a model should be interesting for understanding effects of dissipative disorder on thermalization and transport properties in a quantum many-body system. The master equation then takes the form
\begin{eqnarray}
 &\dot{\rho}=-i[H,\rho]+\sum_{l}(2L^{\dag}_l{\rho}L_l-\{L^{\dag}_lL_l{\rho}\})\nonumber,\\
 &L_l=\sum_{i,j}M_l^{ij}(\sigma_i^x\sigma_{j}^x+\sigma_i^y\sigma_{j}^y+\sigma_i^z\sigma_{j}^z),
 \label{Eq:RandDiss}
\end{eqnarray}
where each \(M_l\) is an arbitrary complex matrix with \(i,j\) indexing its elements. The jump operators $L_{l}$ describe dissipation proportional to the dipole interaction between sites \(i\) and \(j\), with \(M_l^{ij}\) quantifying the strength of this interaction. Whilst not fully realistic, this type of dissipation is instructive to our theory. The Lindblad operator commutes with all generators of the SU(2) symmetry and thus we can identify a series of non-Abelian strong symmetries via the spin operators
\begin{equation}
 [L_l,S^x]=[L_l,S^y]=[L_l,S^z]=0 ,\ \forall l.    
\end{equation}

\begin{figure}[tb]
\begin{indented}
   \item \includegraphics[width=0.8\columnwidth]{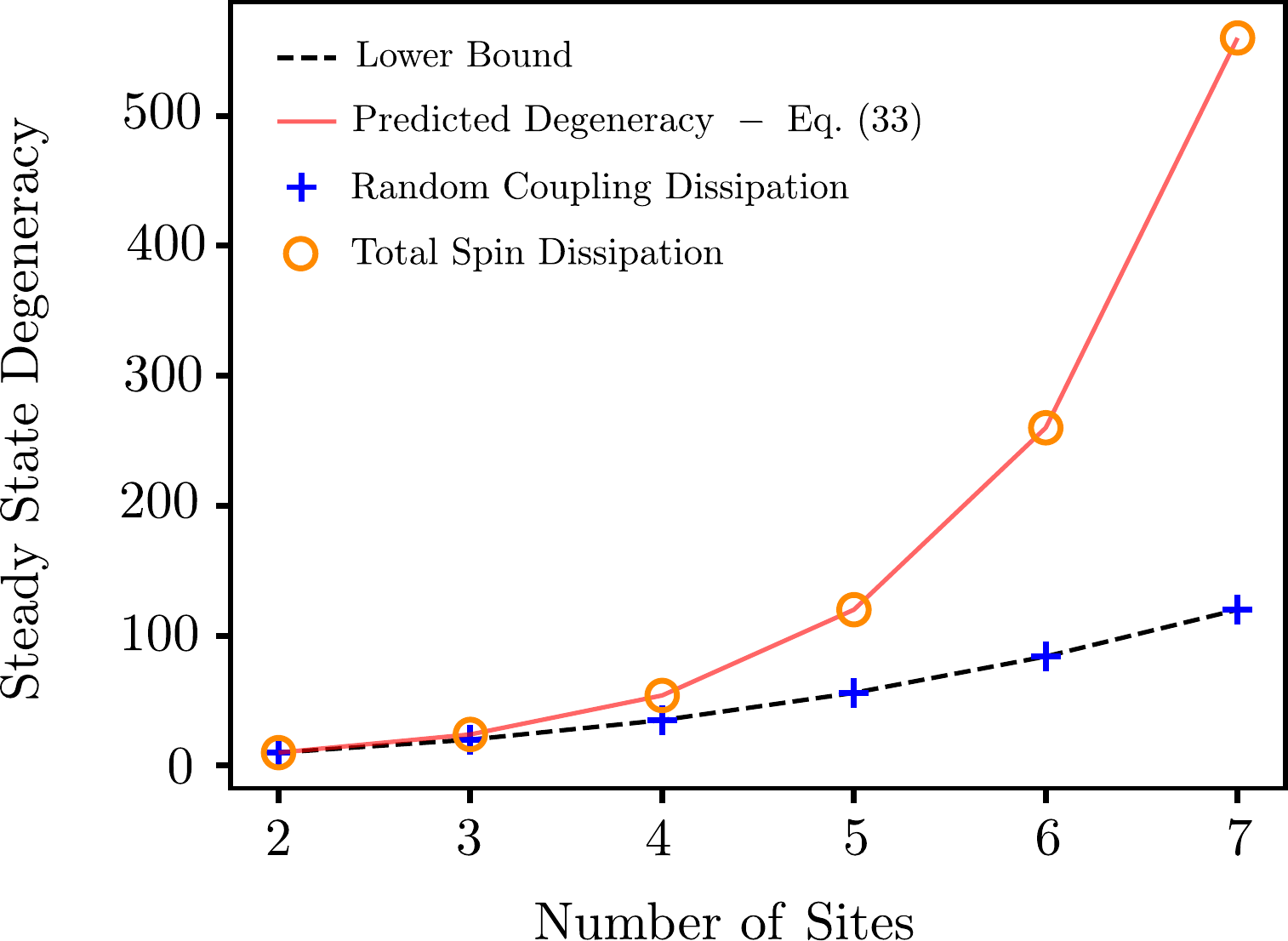}
 \end{indented}
 \caption{\label{fig:ssdxxx} Plot of stationary state degeneracy vs system size for the open XXX Heisenberg model with dissipation which preserves the SU(2) symmetry structure. Two different types of dissipation are considered, `Total spin dissipation' and `Arbitrary coupling dissipation' described in equations~(\ref{Eq:RandDiss}) and (\ref{Eq:TotalDiss}), respectively.
 Data points are calculated by exact diagonalization of the Liouvillian of the system.
 The dashed black line is the theoretical lower bound calculated using
  Theorem ~\ref{theorem} (see table~\ref{table:XXX}),
 while the dashed line is the prediction from equation ~\eqref{eq:casimir-dissipation-bound} which accounts for the additional structure of the nullspace of the system in the case of total spin dissipation.
 }
\end{figure}

We test our lower bound by computing the stationary state degeneracy numerically via exact diagonalization.
 We generate the $M_l^{ij}$ by drawing the real and imaginary parts as random numbers with a uniform distribution over the interval $[-1,1]$.
We then average the stationary state degeneracy over a series of instances of this dissipation.
As shown in figure~\ref{fig:ssdxxx}, we find that our bound is saturated and provides an exact description of the stationary state degeneracy of the system.

\subsubsection{Total spin dissipation.\label{ssec:total-spin-dissip}}
As a second example of dissipation which preserves the SU(2) symmetry, we consider dissipation induced by the Casimir operator:
\begin{eqnarray}
 L & =S^2=(S^x)^2+(S^y)^2+(S^z)^2 ,
 \label{Eq:TotalDiss}
\end{eqnarray}
such that the master equation reads
\begin{eqnarray}
 \dot{\rho} &= -\im [H,\rho]
 +\gamma (2L{\rho}L^{\dag} - \{L^{\dag}L, {\rho}\}).
\end{eqnarray}
Since \([S^2,S^x]=[S^2,S^y]=[S^2,S^z]=0\), the SU(2) symmetry is not broken by the dissipation and thus, again, we have a set of non-Abelian strong symmetries. 

Again, we compute the actual stationary state degeneracy numerically and compare to our theory in table~\ref{table:XXX}.
We report the results in figure~\ref{fig:ssdxxx}. They indicate that for systems with site number $N > 2$, the actual number of stationary states is much greater than the lower bound.
We can, however, explain this difference within the structure of our theory. 
As we will discuss now, there is a further structure to the states within each irreducible representation (see Remark 1 in Theorem~\ref{theorem}) which, consequently, increases the steady state degeneracy above the lower bound.  


Since \(L=S^2\) commutes with the Hamiltonian and $L$ is Hermitian there is a basis in which both $H$ and $L$ are mutually diagonal. Clearly, $\LL$ is also diagonal in this basis. We can write the Casimir operator as, 
\begin{equation}
L=\sum_{i=1}^s \sum_{j=1}^{D_i} \sum_{x=1}^{M_i} c_i \ket{i,j,x}\bra{i,j,x}, 
\end{equation}
where $j$ labels states within each irreducible representation and $x$ labels the \(M_i\) further states in the $\ket{i,j}$ subspace, which have well-defined total angular momentum and angular momentum in the \(z\) direction. For example, $M_{1} = 1, M_{2} = 4$ and $M_{3} = 5$ for the 5-site system whose decomposition is shown in table \ref{table:XXX}. We note that $L$ is the identity within each $\ket{i,j}$ subspace.
\par By Schur's lemma $H$ is proportional to the identity matrix in each subspace $i$,
\begin{equation}
H=\sum_{i=1}^s \sum_{j=1}^{D_i} \sum_{x=1}^{M_i}\sum_{y=1}^{M_i} h_{i}(x,y) \ket{i,j,x}\bra{i,j,y}. 
\end{equation}
 We can then diagonalize $H$ in the $\ket{i,j}$ subspace, 
\begin{equation}
H = \sum_{i=1}^s \sum_{j=1}^{D_i} \sum_{x=1}^{M_{i}}  E^{(i)}_{x} \ket{i,j,x}\bra{i,j,x},
\label{Eq:HeiEig}
\end{equation}
where $E^{(i)}_x$ are the energies within the irreducible representation $i$. Note that there may be further accidental degeneracies in the energies, or more generally additional symmetry structure guaranteeing that for some energies $E^{(i)}_x=E^{(i)}_{x'}$.   

It is clear that $\ket{i,j,x}$ is an eigenstate of both $H$ and $L$ with the same respective eigenvalues $\forall j$ and every pair $x,x'$ that has the same energy $E^{(i)}_x=E^{(i)}_{x'}$ . Therefore, 
\begin{equation}
\LL \ket{i,j,x}\bra{i,j',x'}=0, \qquad \forall x,x' \mid E^{(i)}_x=E^{(i)}_{x'}.
\end{equation}
Hence, assuming that the energies $E^{(i)}_x$ are all distinct the stationary state degeneracy is,
\begin{equation}
  N_s=\sum_{i=1}^s M_i\times D_i^2
  \label{eq:casimir-dissipation-bound}
\end{equation}
which is larger than the lower bound of Theorem ~\ref{theorem} if one takes into account only the strong $SU(2)$ symmetry and matches the exact-diagonalization results, see figure~\ref{fig:ssdxxx}.


\subsection{Hubbard model}

As a final example we consider the Hubbard model under local dissipation. The Hubbard model is a simple but very successful physical model of solid state systems under the tight-binding approximation. For sake of simplicity we focus on the one-dimensional case. The Hamiltonian for an $N$-site lattice reads
\begin{eqnarray}
 H=
 -t\sum_{\langle ij \rangle, \sigma} \left( c_{i,\sigma}^{\dag}c_{j,\sigma} + H.c. \right)
 + U\sum_{i=1}^N n_{i,\uparrow}n_{i,\downarrow},
 \label{Eq: FHModel}
\end{eqnarray}
where $c_{i,\sigma}$ and its adjoint are the annihilation and creation operators for a fermion of spin $\sigma \in \{\uparrow, \downarrow\}$ on site $i$. In addition, $n_{i,\sigma}$ is the number operator for a spin $\sigma$ on site $i$. The quantities $t$ and $U$ are, respectively, kinetic and interaction energy scales, with the summation in the kinetic term taken over nearest-neighbour pairs $\langle ij \rangle$ on the lattice. 

The Hubbard model has a rich symmetry structure which has made it a natural environment for inducing exotic non-equilibrium phases through dissipation \cite{Buca2019, 1902.05012}. The continuous symmetry of this model is SO(4) and because ${\rm SO(4)=SU(2)\times SU(2)/Z_2 }$ there are two commuting SU(2) symmetries within the model~\cite{1dhubbard}. 

The first of these is the spin symmetry, generated by the spin operators
\begin{eqnarray}
 S^{+}=\sum_{j=1}^{N} &&c_{j,\uparrow}^{\dag}c_{j,\downarrow}, \quad S^{-}=\sum_{j=1}^{N} c_{j,\downarrow}^{\dag}c_{j,\uparrow}, 
 \quad 
 S^z=\frac{1}{2}\sum_{j=1}^N(n_{j,\uparrow}-n_{j,\downarrow}),
\end{eqnarray}
which satisfy the SU(2) relations
\begin{eqnarray}
 [S^z,S^{\pm}&&]=\pm S^{\pm},\quad [S^+,S^-]=2S^z,
 \quad
 [H,S^z]=[H,S^\pm]=0.
\end{eqnarray}
The second, hidden, symmetry is often referred to as the ‘\(\eta\)-symmetry’, and relates to spinless quasi-particles (doublons and holons). It plays an important role in the formation of superconducting eigenstates of the Hubbard model \cite{ODLRO}. This $\eta$-symmetry is generated by the operators
\begin{eqnarray}
 \eta^+ = \sum_{i=1}^N(-&&1)^i c_{i,\uparrow}^{\dag}c_{i,\downarrow}^{\dag},
 \quad
 \eta^- = (\eta^+)^{\dag}, 
 \quad
 \eta^z = \frac{1}{2}\sum_{i=1}^N(n_{i,\uparrow}+n_{i,\downarrow}-1),
\end{eqnarray}
which satisfy the relations
\begin{eqnarray}
 [\eta^z,\eta^{\pm}&&]=\pm \eta^{\pm},
 \quad
 [\eta^+,\eta^-]=2\eta^z,
 \quad
 [H,\eta^z]=[H,\eta^\pm]=0.
\end{eqnarray}
The two SU(2) symmetries are abelian:
\[[S^\alpha,\eta^\beta]=0,\quad \alpha,\beta=+,-,z\].

To study the representation of this SU(2)  $\times$ SU(2) symmetry over an $N$-site lattice, we first need to describe its representation in terms of a single site. 
Since the two SU(2) symmetries are commuting, we are able to describe their representation with two individual representations. 
We define a representation of SU(2) $\times$ SU(2) as \((\textbf{m},\textbf{n})\) where \(\textbf{m}\) and \(\textbf{n}\) describing the representations of the first and second SU(2) symmetries respectively.
For instance, we can decompose the single-site SU(2) $\times$ SU(2) group as
\begin{equation}
 (\textbf{2},\textbf{1}\oplus\textbf{1})\oplus(\textbf{1}\oplus\textbf{1},\textbf{2}).
 \label{eq:2su2}
\end{equation}
This means that the 4-dimensional single-site Hilbert space is a direct sum of two invariant subspaces. In the first subspace the representation of the spin symmetry is two dimensional and the representation of \(\eta\) symmetry is the direct sum of two trivial representations, while in the second subspace the converse is true. This can be seen by writing down the spin and \(\eta\) operators on the one-site Hilbert space in explicit matrix form:
\begin{eqnarray}
 S^x &=\frac{1}{2}
 \left[ \begin{array}{cccc}
   0 & 0 & 0 & 0\\
   0 & 0 & 0 & 0\\
   0 & 0 & 0 & 1\\
   0 & 0 & 1 & 0
   \end{array} \right], \quad
 S^z &=\frac{1}{2}
 \left[ \begin{array}{cccc}
   0 & 0 & 0 & 0\\
   0 & 0 & 0 & 0\\
   0 & 0 & -1 & 0\\
   0 & 0 & 0 & 1
   \end{array} \right],
\\
 \eta^x &=\frac{1}{2}
 \left[ \begin{array}{cccc}
   0 & 1 & 0 & 0\\
   1 & 0 & 0 & 0\\
   0 & 0 & 0 & 0\\
   0 & 0 & 0 & 0
   \end{array} \right], \quad
 \eta^z &=\frac{1}{2}
 \left[ \begin{array}{cccc}
   -1 & 0 & 0 & 0\\
   0 & 1 & 0 & 0\\
   0 & 0 & 0 & 0\\
   0 & 0 & 0 & 0
   \end{array} \right].
\end{eqnarray}
The basis of the above matrices is \(\{ {\mid} {\rm vac} \rangle,{\mid}\uparrow \downarrow\rangle,{\mid} \downarrow\rangle,{\mid} \uparrow\rangle \}\), where the arrows indicate the presence of a fermion of either spin `up' or `down'.

We now couple the Hubbard Hamitonian in equation~(\ref{Eq: FHModel}) to an environment which induces homogeneous, local spin-dephasing. The dynamics is modelled, under the Markov approximation, via the Lindblad master equation
\begin{equation*}
 \dot\rho
 =-\im [H,\rho]
 +\gamma \sum_{l=1}^N(2L_l{\rho}L^{\dag}_l
  -\{L^{\dag}_lL_l{\rho}\}),
\end{equation*}
with the Lindblad operators
\begin{equation}
 L_l=s^z_l=\frac{1}{2}(n_{l,\uparrow}-n_{l,\downarrow}).
\end{equation}
A possible experimental implementation for this master equation is discussed in Ref.~\cite{1902.05012} where the spin dephasing was shown to induce superconducting order in the long-time limit of the system.
Under this dephasing, the \(\eta\) symmetry is preserved $[L_l,\eta^{+,-, z}]=0$ whilst the spin SU(2) symmetry is broken into a U(1) symmetry since \([L_l,S^{\pm}]\neq 0\) and \([L_l,S^z]=0\). 

We represent this SU(2) $\times$ U(1) symmetry over the one-site space as
\begin{equation}
 \textbf{2}_0\oplus\textbf{1}_1\oplus\textbf{1}_{-1},
\end{equation}
where the bold numbers denote the representation of the remaining SU(2) symmetry, and the subscript indexes the value of $S^z$ (the U(1) symmetry) in that representation. 
With this notation, we can then construct the representation over the full Hilbert space for $N$ sites. For example, the representation of the 2-site system is
\begin{eqnarray}
    (\textbf{2}_0 & \oplus\textbf{1}_1\oplus\textbf{1}_{-1})
    \otimes
    (\textbf{2}_0\oplus\textbf{1}_1\oplus\textbf{1}_{-1})
    = \nonumber\\
    & \textbf{3}_0\oplus\textbf{1}_0\oplus\textbf{2}_1
    \oplus\textbf{2}_{-1}\oplus\textbf{2}_1
    \oplus\textbf{1}_2\oplus\textbf{1}_0 
    \oplus\textbf{2}_{-1}\oplus\textbf{1}_0\oplus\textbf{1}_{-2},
\end{eqnarray}
which can immediately be extended to $N$ sites.
Using this representation we can apply Th.~\ref{theorem} and predict the lower bound of the stationary space dimension. 
We find that the lower bound of the null space dimension for an \(N\) site system is
\begin{equation}
 N_s=\sum_{n=1}^{N+1}(N+2-n)n^2.  
\end{equation}
In table~\ref{table:Hubbard}, we compare this lower bound with numerical exact-diagonalization results. We find that the
bound is exactly saturated 
for the small systems where calculations are accessible; larger systems are outside numerical tractability.
We notice that, for this example, the growth of this bound is \(O(N^4)\), which is faster than the usual \(O(N^3)\) growth because of the additional \(U(1)\) symmetry.

\begin{table}
 \caption{\label{table:Hubbard}
 The symmetric reduction and corresponding lower bound of the stationary state degeneracy for the spin-dephased Fermi-Hubbard model, which is SU(2) $\times$ U(1) symmetric. The second column lists the unique fundamental representations of this SU(2) $\times$ U(1) symmetry over an $N$-site lattice. The third column is the lower bound on the null space dimension, calculated using Theorem ~\ref{theorem}. The fourth column is the actual stationary state degeneracy obtained from numerically diagonalizing the Liouvillian of the system.}
 \begin{indented}
 \item \begin{tabular}{cccc} 
 \br 
 $N$ & Irreducible representations & Degeneracy lower bound & Actual degeneracy 
 \\ \mr 
 2
   & \begin{tabular}[c]{@{}c@{}}\ $\textbf{3}_0$ \\ $\textbf{2}_1\quad \textbf{2}_{-1}$ \\ $\textbf{1}_2\quad \textbf{1}_0\quad \textbf{1}_{-2}$ \end{tabular}
   & 20  &20 \\ \mr 
 3
   & \begin{tabular}[c]{@{}c@{}} $\textbf{4}_0$ \\ $\textbf{3}_1\quad \textbf{3}_{-1}$ \\ $\textbf{2}_2\quad \textbf{2}_0\quad \textbf{2}_{-2}$ \\ $\textbf{1}_3\quad \textbf{1}_1\quad \textbf{1}_{-1}\quad \textbf{1}_{-3}$\end{tabular}
   & 50  &50 \\ \mr 
 4
   & \begin{tabular}[c]{@{}c@{}}$\textbf{5}_0$ \\ $\textbf{4}_1\quad \textbf{4}_{-1}$\\ $\textbf{3}_2\quad \textbf{3}_0\quad \textbf{3}_{-2}$ \\ $\textbf{2}_3\quad \textbf{2}_1\quad \textbf{2}_{-1}\quad \textbf{2}_{-3}$ \\ $\textbf{1}_4\quad \textbf{1}_2\quad\textbf{1}_0\quad \textbf{1}_{-2}\quad \textbf{1}_{-4}$\end{tabular}
   & 105 & 105 \\ \mr 
 5
   & \begin{tabular}[c]{@{}c@{}} $\ \textbf{6}_0 $\\$ \textbf{5}_1\quad \textbf{5}_{-1} $ \\ $\textbf{4}_2\quad \textbf{4}_{0}\quad \textbf{4}_{-2}$\\ $\textbf{3}_3\quad \textbf{3}_1\quad \textbf{3}_{-1}\quad \textbf{3}_{-3}$ \\ $\textbf{2}_4\quad \textbf{2}_2\quad \textbf{2}_{0}\quad \textbf{2}_{-4}\quad \textbf{2}_{-2}$ \\ $\textbf{1}_5\quad \textbf{1}_3\quad\textbf{1}_1 $ \\ $ \textbf{1}_{-1}\quad \textbf{1}_{-3}\quad \textbf{1}_{-5}$ \end{tabular}
   & 196  & 196 
   \\ \br 
 \end{tabular}
 \end{indented}
\end{table}

\section{Conclusion}
We have provided a lower bound on the stationary-state degeneracy of open quantum systems with multiple non-Abelian symmetries. This bound is based on the decomposition of the symmetry group into a series of irreducible representations. As examples we have studied an open quantum network, the dephased XXX spin chain and a spin-dephased Hubbard model.
By comparing with exact-diagonalization calculations, we have shown that the bound is saturated in most of the examples we looked at, providing crucial evidence of the strength of our bound for open quantum many-body systems. Thus we expect that if additional degeneracies are discovered in the stationary state through e.g. numerical calculations, these may hint at possible hidden symmetries in the model waiting to be discovered. Alternatively, the system may have accidental degeneracies. 

Our general results here open many interesting venues for future study. The most pressing is extending the results to non-Abelian dynamical symmetries. Dynamical symmetries in both closed \cite{Marko}, and open quantum many-body systems \cite{Buca2019, Tindall2019,BucaJaksch} have been shown to guarantee absence of any relaxation to stationarity, instead leading to persistent oscillations and complex dynamics. Our work should also be extended beyond the Markovian approximation, to include more widely applicable Redfield master equations \cite{openbook}.
In cases where the open system has only approximate non-Abelian strong symmetries, one may expect metastable stationary states~\cite{metastability,metastability2}.

This work also opens up the possibility of applying our symmetry reduction method to the study of dissipative magnetic systems \cite{Manzano3}, and other two-dimensional systems \cite{ChiralBICs,vanHoveBICs}, where the high level of symmetry reduction could be beneficial due to a lack of viable computational methods. 
We also anticipate further work applying our theory to dissipation-induced frustration of open quantum systems~\cite{introtofrust} due to the high entropy content of the stationary state.

Lastly, our work on the fully-connected quantum network provides a systematic method to study more complex networks with hierarchical structure and topology \cite{HurtadoManzano2}, as well as other molecular structures with a high degree of symmetry~\cite{Jordi3,Thingna2016}.

\ack 
We would like to thank R. A. Molina for useful discussions.
This work has been supported by UK EPSRC grants Nos.\  EP/P01058X/1, 
EP/P009565/1, 
EP/K038311/1, 
EPSRC National Quantum Technology Hub in Networked Quantum Information Technology (EP/M013243/1), and the European Research Council under the European Union's Seventh Framework Programme (FP7/2007-2013)/ERC Grant Agreement No.\ 319286 (Q-MAC).   
Zh.\ Zhang acknowledges support from the Tsinghua University Education Foundation (TUEF).


\section*{References}

\bibliographystyle{unsrt}
\bibliography{zhao} 

\end{document}